\documentclass[aps,%
pre,%
aps,%
twocolumn,%
showpacs,%
]{revtex4-1}

\usepackage{amsmath}
\usepackage{amssymb}
\usepackage{graphicx}

\def \w {\omega}

\begin{document}

\title{Synchronization of a Josephson junction array in terms of global variables}

\author{Vladimir Vlasov}
\email{mr.voov@gmail.com}
\author{ Arkady Pikovsky}
\affiliation{Department of Physics and Astronomy, Potsdam University, 14476
Potsdam, Germany}

\begin{abstract}
We consider an array of Josephson junctions with a common LCR-load. Application of the
Watanabe-Strogatz approach [Physica D, v. 74, p. 197 (1994)] allows us to formulate the 
dynamics of the array via the global
variables only. For identical junctions this is a finite set of equations, analysis of
which reveals the regions of bistability of the 
synchronous and asynchronous states. For disordered arrays with
distributed parameters of the junctions, 
the problem is formulated as an integro-differential
equation for the global variables, here stability of the asynchronous states and the 
properties
of the transition synchrony-asynchrony are established numerically.  
\end{abstract}

\date{\today}
\maketitle

\section{Introduction}
\label{sec:intro}

Synchronization in populations of coupled oscillators is a general phenomenon
observed in many physical systems, see recent experimental studies of optomechanical, 
micromechanical, electronic, mechanical, chemical 
oscillators ~\cite{Heinrich_etal-11,*Zhang_etal-12,*Temirbayev_etal-12,%
*Martens_etal-13,*Tinsley_etal-12}. Synchronization effects are also ubiquitous
in biology and social sciences. One of the basic examples  of oscillating physical 
systems that being coupled synchronize, are Josephson 
junctions~\cite{Jain84,*Benz-Burroughs-91,%
*Whan-Cawthorne-Lobb-96,*Cawthorne_etal-99}. In theoretical studies of the Josephson 
junction arrays one typically either performs direct
numerical simulation of the microscopic equations  
(see, e.g., \cite{Hadley-Beasley-Wiesenfeld-88,*Filatrella_etall-00}) 
or reduces the problem to the standard Kuramoto-type 
model~\cite{Wiesenfeld-Swift-95,Wiesenfeld-Colet-Strogatz-96,Wiesenfeld-Colet-Strogatz-98}. 

Quite remarkable in this respect is the
paper~\cite{Heath-Wiesenfeld-00}, where a careful comparison of the microscopic
modeling and the reduced Kuramoto-type model has been performed. The authors 
demonstrated that 
a hysteretic transition to synchrony in an array of
Josephson junctions can be explained by a Kuramoto-type modeling (where usually 
the transition is not hysteretic), if in its derivation one self-consistently
accounts for changes of the oscillator parameters. 

Our aim in this paper is to
shed light on the hysteretic transitions to synchrony in Josephson arrays by 
studying the equations for global variables. In this approach, that is based
on the seminal papers by Watanabe and 
Strogatz (WS)~\cite{Watanabe-Strogatz-93,Watanabe-Strogatz-94}, it is possible to formulate
exact low-dimensional equations for the array, without using approximate reduction to the Kuramoto model.  
The paper is organized as follows. First, we formulate the equations for the array of \textit{identical} 
junctions via the global variables. Analysis of these equations shows regions of bistability asynchrony--synchrony, 
and the hysteretic transitions. Then we proceed to
\textit{non-identical} junctions, where the equations are of more complex form.
Here we analyze stability of asynchronous states, and show numerically
that the transition
to synchrony is also hysteretic.

\section{Identical Junctions} 
\label{sec:ij}

\subsection{Formulation in terms of global variables}

We start with formulating the system of equations for the Josephson junction series array with a LCR load. 
Our setup is the same as in refs.~\cite{Wiesenfeld-Swift-95,Wiesenfeld-Colet-Strogatz-96,Wiesenfeld-Colet-Strogatz-98}, 
the equatons for the junction phases $\varphi_i$ and the load capacitor charge $Q$ read
\begin{equation}
\begin{aligned}
{\hbar\over 2er}\frac{d\varphi_i}{dt}+I_c\sin\varphi_i&=I-\frac{dQ}{dt}\;,\\
L\frac{d^2Q}{dt^2}+R\frac{dQ}{dt}+\frac{Q}{C}&={\hbar\over 2e}\sum_{i=1}^N \frac{d\varphi_i}{dt}\;.
\end{aligned}
\label{eq:beq1}
\end{equation} 
Here $N$ is the number of junctions, described by a resistive model with critical current $I_c$ 
and resistance $r$, while $L,C,R$ are parameters of the LCR-load. It is convenient to introduce dimensionless 
variables according to
\begin{equation}
		\label{eq:dimens}
		\begin{gathered}
			\w_c = 2erI_c / \hbar, \ \ t^*=\w_c t, \\ 
			 \ \ Q^*=\w_c L^* Q/ I_c, \ \ I^*=I/I_c, \\
			R^*=R/rN,\ \ L^*=\w_c L/rN,\ \ C^*=N\w_c rC\;,
		\end{gathered}
\end{equation}
and to rewrite the system (\ref{eq:beq1}) in a dimensionless form (droping asterixes for simplicity)
	\begin{equation}
\begin{aligned}
	\dot{\varphi_i}&=I-\epsilon\dot{Q} -\sin\varphi_i,\\
		\ddot{Q} + \gamma\dot{Q}+ \w_0^2 Q&=I-{1\over N}\sum_{i=1}^N \sin\varphi_i,
\end{aligned}
\label{eq:beq2}
\end{equation}
where $\epsilon=1/L^*$, $\gamma=(R^*+1)/L^*$, and $\w_0=1/\sqrt{L^*C^*}$.

The global coupling can be represented through the complex mean field (Kuramoto order parameter) 
\begin{equation}
		\label{eq:cmf}
		\begin{aligned}
		Z = r e^{\text{i}\theta} &= {1\over N}\sum_{i=1}^N (\cos\varphi_i + \text{i}\sin\varphi_i)\;,\\
		{\rm Im}(Z)&={1\over N}\sum_{i=1}^N \sin\varphi_i\;,
		\end{aligned}
\end{equation}
and the equations for the junction phases can be written as
\begin{equation}
		\label{eq:pheq}
		\dot{\varphi_i}=I-\epsilon\dot{Q} +{\rm Im}(e^{-\text{i}\varphi_i}).
\end{equation}
This form of the phase equation allows us to use the Watanabe-Strogatz 
ansatz~\cite{Watanabe-Strogatz-93,Watanabe-Strogatz-94}, applicable to general
systems of phase equations driven by a common force and having form
\begin{equation}
		\label{eq:pheqws}
		\dot{\varphi_i}=f(t) +{\rm Im}(G(t)e^{-\text{i}\varphi_i})
\end{equation}
with arbitrary real $f(t)$ and complex $G(t)$ (in our case $f=I-\epsilon \dot{Q}$,
$G=1$). We use the formulation of the Watanabe-Strogatz theory 
presented in Ref.~\cite{Pikovsky-Rosenblum-11}. 
The ensemble is characterized by three global time-dependent WS
variables $\rho,\Phi,\Psi$ 
and $N$ constants of motion $\psi_i$ (of which only $N-3$ are independent)
which are related to the phases $\varphi_i$ as
 \begin{equation}
		\label{eq:wstrans}
	e^{\text{i}\varphi_i}=e^{\text{i}\Phi}\frac{\rho+\exp(\text{i}(\psi_i-\Psi))}{\rho\exp(\text{i}(\psi_i-\Psi))+1}
\end{equation}
with additional conditions $\sum_i \cos\psi_i=\sum_i\sin\psi_i=\sum_i\cos 2\psi_i=0$. The equations for the global WS 
variables read~\cite{Watanabe-Strogatz-93,Watanabe-Strogatz-94,Pikovsky-Rosenblum-11}
\begin{equation}
		\label{9}
		\begin{aligned}
			\dot{\rho}&=\frac{1-\rho^2}{2}{\rm Re}(e^{-\text{i}\Phi}), \\
			\dot{\Psi}&=\frac{1-\rho^2}{2\rho}{\rm Im}(e^{-\text{i}\Phi}), \\
			\dot{\Phi}&=I-\epsilon\dot{Q}+\frac{1+\rho^2}{2\rho}{\rm Im}(e^{-\text{i}\Phi}).
		\end{aligned}
\end{equation}
To close the system we need to add the equation for $Q$, where the imaginary part of the 
order parameter $Z$ enters, so $Z$ should be represented 
through the Watanabe-Strogatz variables. In general, the expression for $Z$ is rather complex
 (cf.~\cite{Pikovsky-Rosenblum-08,Pikovsky-Rosenblum-11}) but in the case of a uniform distribution of 
the constants $\psi_i$,
the order parameter is just $Z=\rho e^{\text{i}\Phi}$. 
This important case, where WS global variables $\rho,\Phi$ have a
clear physical meaning as the components
of the Kuramoto order parameter, will be treated below.
Additionally, we notice that the variable $\Psi$ does not enter other equations, so we obtain a closed system of equations that describes the array
\begin{equation}
 		\label{eq:sysws}
		\begin{aligned}
			\dot{Z}&=\text{i}(I-\epsilon\dot{Q})Z+{1\over 2}-{Z^2 \over 2}, \\
			\ddot{Q} + \gamma\dot{Q}+\w_0^2 Q&= I-{\rm Im}(Z).
		\end{aligned}
\end{equation}

\subsection{Bistability and hysteretic transitions}

Analysis of system (\ref{eq:sysws}) is our goal in the rest of this section. Before proceeding, some remarks are in order.
First, in the derivation of (\ref{eq:sysws}) no approximation except for an assumption of a uniform distribution 
of constants $\psi_i$, has been made. The latter is a restriction on initial conditions, we discuss its relevance below. Second, 
the order parameter $Z$ does not vanish in the case of full asynchrony of junctions: for noncoupled 
junctions with $\epsilon=0$
we get a steady state $Z_0=\text{i}(I-\sqrt{I^2-1})$. This non-vanishing value appears because free junctions 
rotate non-uniformly 
and the  ``natural'' distribution of the phases in the asynchronous state is not uniform.

We start the analysis of (\ref{eq:sysws}) with finding its steady states. Because at such a state $\dot Q=0$, the coupling 
vanishes and the steady state describing the asynchronous regime with  $Z_0=\text{i}(I-\sqrt{I^2-1})$, 
$Q_0=\w_0^{-2}\sqrt{I^2-1}$ is the only stationary solution. Stability of this solution is determined by the fourth-order
characteristic equation
	\begin{equation}
		\label{eq:chareq}
		\begin{aligned}
		\lambda^4 + \gamma \lambda^3 &+ (\w_0^2+I^2-1)\lambda^2 + \\
		+[(\gamma - \epsilon)(I^2-1)&+\epsilon I \sqrt{I^2-1}\;]\lambda + \w_0^2 (I^2-1) = 0.
		\end{aligned}
	\end{equation}
The stability border can be easily found by assuming $\lambda=\text{i}\w$:
\begin{equation}
 		\w_0^2=(I^2-1)+{\epsilon \over \gamma}\sqrt{I^2-1}(I-\sqrt{I^2-1}).
\label{eq:stbord}
\end{equation}

\begin{figure}
		\includegraphics[width=0.8\columnwidth]{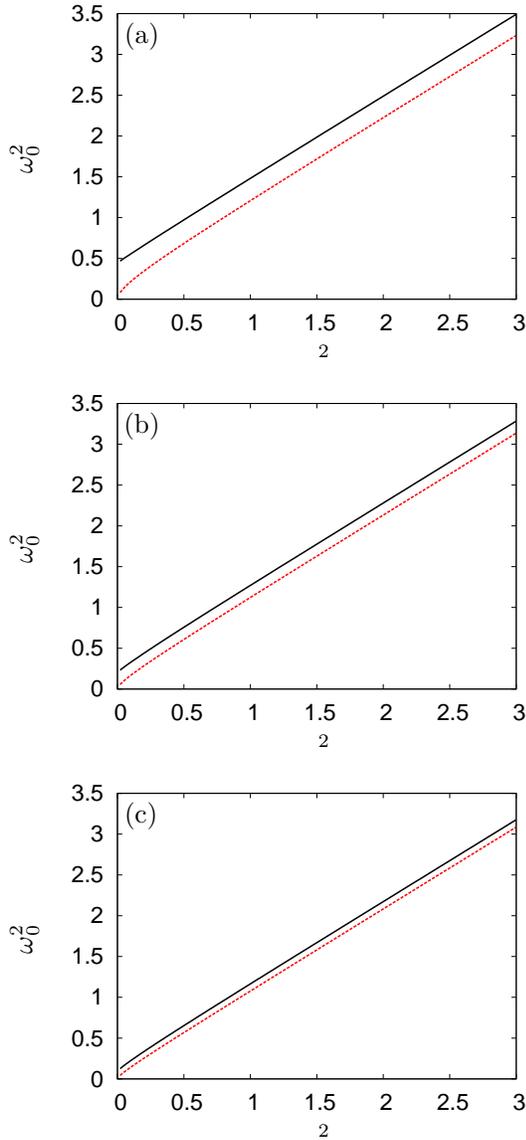}
		\caption{(Color online) Domains of stability
of synchronous (above lower dashed line) and asynchronous (below upper solid line) states on the plane of parameters ($\w_0^2,\Omega^2$), 
where  $\Omega=\sqrt{I^2-1}$ is the natural frequency of the junctions.
Here $\epsilon=0.5$, and  $\gamma=1.0$ (a), $\;1.7$ (b), $\;2.7$ (c). }
		\label{fig:bist}
	\end{figure}

The fully synchronous solution of  (\ref{eq:sysws}) corresponds to the case $|Z|=1$, so that only the phase $\Phi$ 
changes, according to the system 
\begin{equation}
 \begin{aligned}
			\ddot{Q}&+\gamma \dot Q+\w_0^2 Q=I- \sin\Phi\;, \\
			\dot{\Phi}&=I-\epsilon \dot Q-\sin\Phi\;.
 \end{aligned}
\label{eq:lc}
\end{equation}
We have found the limit cycle in Eq.~(\ref{eq:lc}) numerically and determined its stability by finding the largest multiplier.
Together with expression (\ref{eq:stbord}) this allows us to find the domains of stability of the asynchronous and 
synchronous states, together with the region of bistability of these regimes, see Fig.~\ref{fig:bist}.

\begin{figure}
		\includegraphics{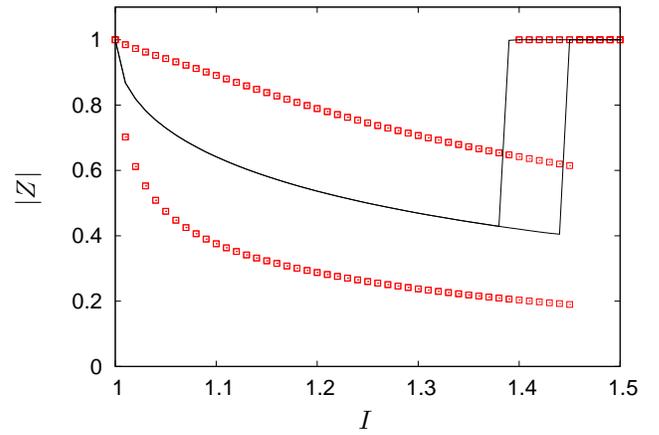}
		\caption{Dependence of the order parameter $|Z|$ on the current $I$
		for 100 junctions. Line: uniform distribution of constants $\psi_i$, squares:
		nonuniform distributions.}
		\label{fig:bist1}
	\end{figure}

In Fig.~\ref{fig:bist1} we give another illustration of the bistability,
presenting the dependence of $Z_0$ on parameter $I$, together with the value $|Z|=1$ in the synchronous case. Here we also show what happens if 
our basic assumption at derivation of eqs.~(\ref{eq:sysws}), namely of a uniform distribution of constants $\psi_i$, is not satisfied.
We have simulated an ensemble of 100 junctions, preparing the initial conditions with a nonuniform distribution of constants $\psi_i$
as described in ref.~\cite{Pikovsky-Rosenblum-11}, appendix C. Instead of leading to a stable state $Z_0$, the desynchronous population now shows an oscillating variable $Z(t)$,
minima and maxima of which are marked with squares. In the synchronous regime, $|Z|=1$ as before, and the information on the constants $\psi_i$ gets lost as synchrony establishes.

\section{Nonidentical Junctions} 
\label{sec:nij}
\subsection{Formulation of the model}
There are two parameters of individual junctions that can differ: the critical current $I_c$ and the 
resistance $r$ (cf.~\cite{Wiesenfeld-Colet-Strogatz-96,Wiesenfeld-Colet-Strogatz-98}). 
In order to be able to apply the Watanabe-Strogatz approach as above, we will assume that they are organized in groups,
each of the size $P$, and the parameters of all junctions in a group are identical: the critical current is $I_c(1+\xi_k)$ and 
the resistance is $r(1+\eta_k)$, where index $k=1,\ldots ,M$ counts the groups. The total number of junctions is $N=MP$.
In this setup the equations for the junctions read
\begin{equation}
  \begin{aligned}
\dot{\varphi}_{ki}&=(1+\eta_k)[I-\epsilon\dot{Q} -(1+\xi_k)\sin\varphi_{ki}]\\
\ddot{Q} + \gamma\dot{Q}+ \w_0^2 Q&=I-{1\over N}\sum_{k=1}^{M} (1+\eta_k)(1+\xi_k) \sum_{i=1}^{P} \sin\varphi_{ki}.
 \end{aligned}
\label{eq:nij}
\end{equation}
To each group the Watanabe-Strogatz ansatz as described in the previous section can be applied, and
as a result instead of the identical array equations~(\ref{eq:sysws}) we obtain a system
\begin{equation}
\begin{gathered}
 		\ddot{Q} + \gamma\dot{Q}+ \w_0^2 Q=I-\langle(1+\eta_k)(1+\xi_k){\rm Im}(Z_k)\rangle, \\
		\dot{Z_k}=(1+\eta_k)\left(\text{i} (I-\epsilon\dot{Q})Z_k+(1+\xi_k){1-Z_k^2 \over 2}\right),
\end{gathered}
\label{eq:niws}
\end{equation}
where average $\langle\rangle$ is taken over all groups. Starting from
(\ref{eq:niws}) one can easily take a thermodynamic 
limit of an infinite number of groups 
$M\to\infty$, in this limit $Z_k \to Z(\eta,\xi)$. Then (\ref{eq:niws}) reduces to an integro-differential equation that includes the 
distribution function $W(\eta,\xi)$ of disorder 
parameters $\xi,\eta$ (cf.~\cite{Pikovsky-Rosenblum-11}):
\begin{equation}
  \begin{aligned}
 		\ddot{Q} + \gamma\dot{Q}&+ \w_0^2 Q=I-\\
		-\iint \,d\eta\,d\xi\; W(\eta,\xi)\,&(1+\eta)(1+\xi){\rm Im}(Z(\eta,\xi))\;, \\
		\dot{Z}(\eta,\xi)=(1+\eta)&\left(\text{i} (I-\epsilon\dot{Q})Z+(1+\xi){1-Z^2 \over 2}\right).
\end{aligned}
\label{eq:niwsint}
\end{equation}

\subsection{Asynchronous state and its stability}
	\begin{figure}
		\includegraphics{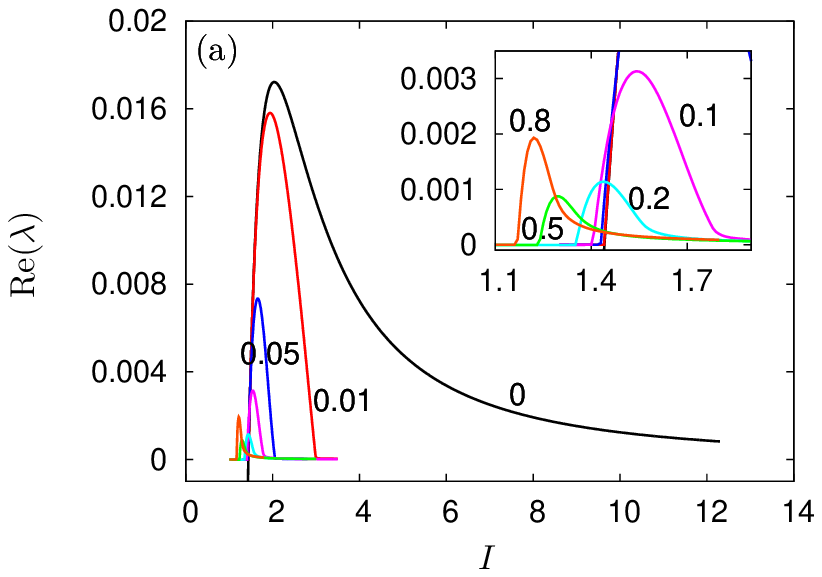}
		\includegraphics{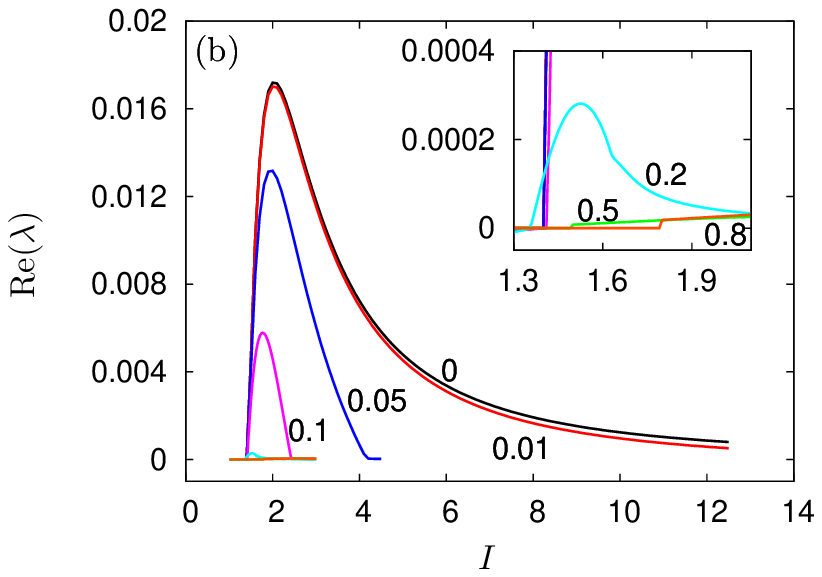}
		\caption{(Color online) Real part of the maximum eigenvalue $\lambda$ as a 
function of the dimensionless current $I$ for the different values (numbers on the panels) of $\mu$ (panel (a)) and $\zeta$ (panel (b)).}
		\label{fig:ev}
	\end{figure}

The asynchronous state is the steady state of the system (\ref{eq:niwsint}):
\begin{equation}
	\label{eq:fpnin}
	\begin{aligned}
	&Z_0(\eta,\xi)=\text{i}\frac{I-\sqrt{I^2-(1+\xi)^2}}{1+\xi},\\
	&Q_0=\w_0^{-2} \iint \,d\eta\,d\xi\; W(\eta,\xi)\,(1+\eta)\sqrt{I^2-(1+\xi)^2}\;,
	\end{aligned}
\end{equation}
where we assume $\langle \xi\rangle=\langle \eta\rangle=0$.
Remarkably, the disorder in the junction resistances (parameter $\eta$) does not influence the value $Z_0$, only the
disorder in critical currents (parameter $\xi$).  However, the stability of this asynchronous state depends on 
distributions of $\eta$ and $\xi$. We consider two cases, with a disorder in one parameter only.

(i) Disorder in resistances. Here we assume that $W(\eta,\xi)=\delta(\xi)W_{\mu}(\eta)$ where $W_\mu$ is a uniform 
distribution in the interval $(-\mu,\mu)$. To study the perturbations in the integral equation (\ref{eq:niwsint}) at 
the steady solution (\ref{eq:fpnin}), we discretized the integral using 500 nodes and found the eigenvalues of the 
resulting matrix. The results
for the maximal eigenvalue are shown in Fig.~\ref{fig:ev}a. One can see that, with increasing the external current 
$I$, the asynchronous state loses stability almost at the same critical value as for identical 
junctions (expression (\ref{eq:stbord})), but
for large values of $I$ the stability is restored. The region of instability decreases for larger disorder $\mu$.

(ii) Disorder in critical currents. Here we assume that $W(\eta,\xi)=\delta(\eta)W_{\zeta}(\xi)$, where $\zeta$ is the width of the uniform distribution.
With the same procedure as in case (i) we found the stability eigenvalues that are 
shown in Fig.~\ref{fig:ev}b. Qualitatively, the pictures look similar: both disorders result in a finite (in therms of 
the external current $I$) region of instability of the asynchronous state.
	\begin{figure}
		\includegraphics{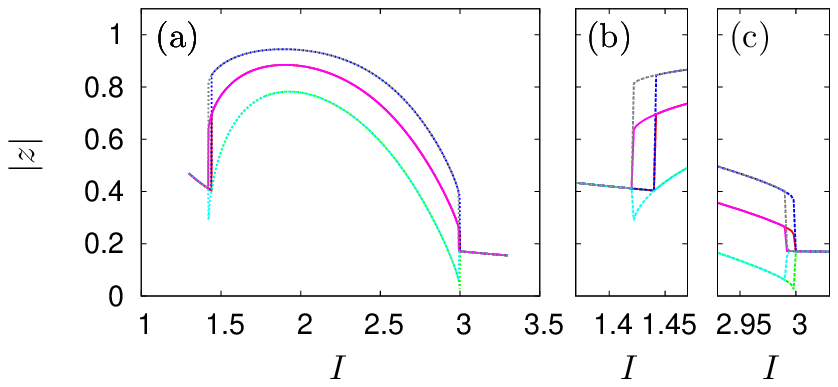} \\
		\includegraphics{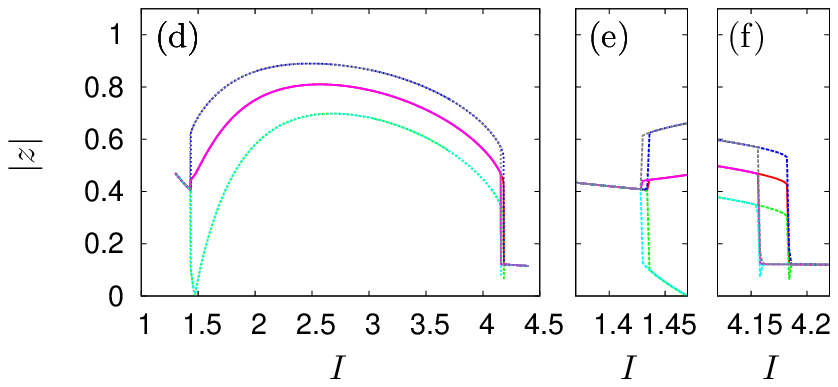}
		\caption{(Color online) Panels (a),(d): Dependence of the averaged order parameter $|z|$ on current $I$, $\mu=0.01,\,
\zeta=0$ and $\mu=0,\,\zeta=0.05$ respectively. Three lines show the maximal (upper dashed line), the 
average (solid line), and the minimal (lower dashed line) value of variations of $|z|$ in time, in the asynchronous states these 
lines coincide. Panels (c), (d), (e) and (f) show enlargements 
of the regions near the synchrony-asynchrony transitions, to demonstrate the hysteresis. }
		\label{fig:evn}
	\end{figure}

Both calculations presented in Fig.~\ref{fig:ev} show, that the main effect of disorder in 
arrays is in the establishing of stability of the asynchronous state
for large values of current $I$, while only in some range (which decreases
with disorder) the asynchrony is unstable. We illustrate the appearing synchrony patterns in 
disordered arrays in the next subsection.

\subsection{Numerical simulations}

Dynamics of the  nonhomogeneous arrays of Josephson junctions is illustrated in Figs.~\ref{fig:evn}. 
As above, we consider not a general situation where both the 
critical current and the resistance are
spread, but cases where one of these parameters has a distribution. 
In numerical simulations we
use the discrete representation (\ref{eq:niws}). In order to avoid spurious 
non-smooth solutions, an additional
very small viscous term $\sim(Z_{k+1}+Z_{k-1}-2Z_k)$ was added 
to the equation for $Z_k$ that ensures 
numerical stabilization of the integro-differential equation.

To characterize synchrony we calculated the average over the array order parameter
$z=M^{-1}\sum_k Z_k$ and plot it vs. parameter $I$ in Fig.~\ref{fig:evn}. In the asynchronous state this 
parameter attains the fixed point (cf. Eq.~(\ref{eq:fpnin})), while in the synchronous state it 
oscillates arround some mean value (because of disorder the synchrony is not complete, so $|z|<1$). 
Remarkably, also in the case of disorder, the transition to synchrony demonstrates
hysteresis both for small and large values of $I$, as can be seen 
on panels (b),(c),(e), and (f) of  Fig.~\ref{fig:evn}.


 \section{Conclusion} 
In this paper we applied the approach by Watanabe and Strogatz to the description of the 
synchronization transition in an array of Josephson junctions with an LCR load. For identical
junctions a closed low-dimensional system of equations for global variables (the Watanabe-Strogatz variables
for the junctions and two variables describing the load) demonstrates a region of bistability
at the transition from asynchrony to full synchrony, so that this transition shows hysteresis. 
This confirms previous results based on the approximate self-consistent reduction to the Kuramoto model~\cite{Heath-Wiesenfeld-00}.
For nonidentical junction the method yields an integro-differential system, as each group of junctions
having certain parameters is described by the WS variables. Here, with the growth of the variability of parameters,
the region of synchronization shrinks. Transition to synchrony in this case is also hysteretic.

Validity of the WS approach to the Josephson junction array is based on the fact, that for standard junctions
the dependence of the superconducting current on the phase is a simple sine function. Therefore, the
theory is also 
valid for so-called $\pi$-junctions~\cite{Buzdin-05}, 
where the current has an opposite direction but nevertheless is propotional
to $\sin(\varphi)$. However, for recently constructed so-called $\varphi$-junctions~\cite{Sickinger_etal-12}, 
where the phase dependence of the
current contains the second harmonics, the WS approach is not applicable, and synchronization
of such junctions remains a challenging problem. 

\begin{acknowledgments}
V. V. thanks the IRTG 1740/TRP 2011/50151-0, funded by the DFG /FAPESP.
\end{acknowledgments}


%

\end{document}